\documentclass{elsart}

\usepackage{amssymb}
\usepackage{latexsym}
\usepackage{graphicx}
\begin{document}

\begin{frontmatter}


\title{Detecting Stochastic Information of Electrocardiograms}
\author{Rafael M. Guti\'{e}rrez,}
\ead{rgutier@uan.edu.co}
\author{ Luis A. Sandoval}
\ead{luisand@uan.edu.co}
\address{Centro de Investigaciones, Universidad Antonio Nari\~{n}o, Bogot\'{a}, Colombia}



\begin{abstract}
In this work we present a method to detect, identify and characterize stochastic information contained in an electrocardiogram (ECG). We assume, as it is well known, that the ECG has information corresponding to many different processes related to the cardiac activity. We analyze scaling and Markov processes properties of the detected stochastic information using the power spectrum of the ECG and the Fokker-Planck equation respectively. The detected stochastic information is then characterized by three measures. First, the slope of the power spectrum in a particular range of frequencies as a scaling parameter. Second, an empirical estimation of the drift and diffusion coefficients of the Fokker-Planck equation through the Kramers-Moyal coefficients which define the evolution of the probability distribution of the detected stochastic information. 
\end{abstract}

\begin{keyword}
Stochastic processes; Electrocardiograms

\end{keyword}

\end{frontmatter}

\section{Introduction}

The analysis of ECGs, or electrocardiography, remains the best non-invasive method for detecting or predicting coronary artery disease, it is also a very practical and nonexpensive method to obtain useful information about the quality of the cardiac activity. In the last few years, new concepts and methods from different fields of physics have been applied to the analysis of ECGs. The ECG can be considered as time series of measurements of an observable, i.e. the electrical activity of the heart recorded from the body surface. This approach has given some interesting results and provided new tools to define new measures of great potential for the characterization of the cardiac activity.  These new measures are able to detect information contained in the ECGs not accessible with traditional methods. Such new useful information may be a mixture of both, stochastic and deterministic information, that has to be unraveled and analyzed with correspondingly different methods. In this work we present a method and some interesting results to analyze ECGs as time series by detecting and characterizing the stochastic information in the context of Markov processes. In section 2 we present the problem of extracting stochastic information from a broad band power spectrum. In section 3 we present the method to characterize the detected stochastic information in the context of Markov processes applied to healthy and pathologic ECGs. Section 5 is devoted to discussion and some conclusions.

\section{Detecting stochastic information in ECGs}

In a previous work we developed a method to separate deterministic and stochastic information from ECGs using the power spectrum, PS, of the ECG \cite{SCI}. The fundamental idea was to improve the $1/f$ behavior of the broad band power spectrum, BBPS, of the ECG in certain range of frequencies. This method uses an archetypal reconstruction of the ECG \cite{CB,Orti}. The difference between the original ECG and the archetypal reconstruction is considered the preliminary stochastic information, PSI. The final stochastic information, SI, was obtained by adjusting the power spectrum of the PSI to the best $1/f^{\alpha}$ power spectrum. The $1/f^{\alpha}$ power spectrum transformed back to the time space was considered the SI, (the difference between the $1/f^{\alpha}$ and the power spectrum of the difference between the original ECG and its archetypal reconstruction, transformed back to the time space, was considered part of the deterministic information). This process was partially successful because the detected SI obtained from the ECGs, showed interesting differences between healthy and pathologic ECGs when analyzed as Markov processes. However, the archetypal reconstruction is not necessarily a method to separate the deterministic and stochastic information, it is a better or worse reconstruction of the whole ECG (depending on the size of the base in use) with its deterministic and stochastic components mixed. The reconstruction of the ECG with a scale dependent base, such as wavelets, complemented by a process of fitting the BBPS to a four parameter function is work in progress to improve the extraction of the SI from the ECG. 
In this work we use some preliminary results of the mentioned work in progress as the SI to be used and differentiate healthy and pathologic cases. Despite the preliminary character of the results used as the SI obtained from 21 healthy and 30 pathologic ECGs \cite{http}, we found important qualitative results of one measure that can distinguish different between the healthy and pathologic cases where other measures do not capture such difference. This measure is the diffusion parameter of the Fokker-Planck equation estimated through the Kramers-Moyal coefficients and the method to estimate it is presented below with some results.

\section{Characterizing stochastic information from ECGs}            

The ECGs have broad band power spectra, BBPS, with complex structure but with approximately $1/f^{\alpha}$ behavior for an important range of the frequency domain. The frequency range where the SI is going to be characterized goes from $\sim 7$Hz to $\sim 20$Hz. Smaller and larger frequencies may be strongly affected by a variety of artifacts related with the different measurement and recording conditions for each ECG associated with nonstationarity and noise contaminations respectively. Indeed, these two regions of the PS are more correlated to the data base used than to the healthy or pathologic character of the ECGs. The estimated value of $\alpha$ is a first characteristic of the SI obtained from the ECG. The different values of $\alpha$ estimated from all the SI obtained from each ECG do not permit us to distinguish between healthy and pathologic cases considering the mentioned range of frequencies. The following procedure will give us two more measures to characterize the SI obtained from from ECGs. 

\begin{figure}
\begin{center}
\includegraphics[height=9cm,width=8cm]{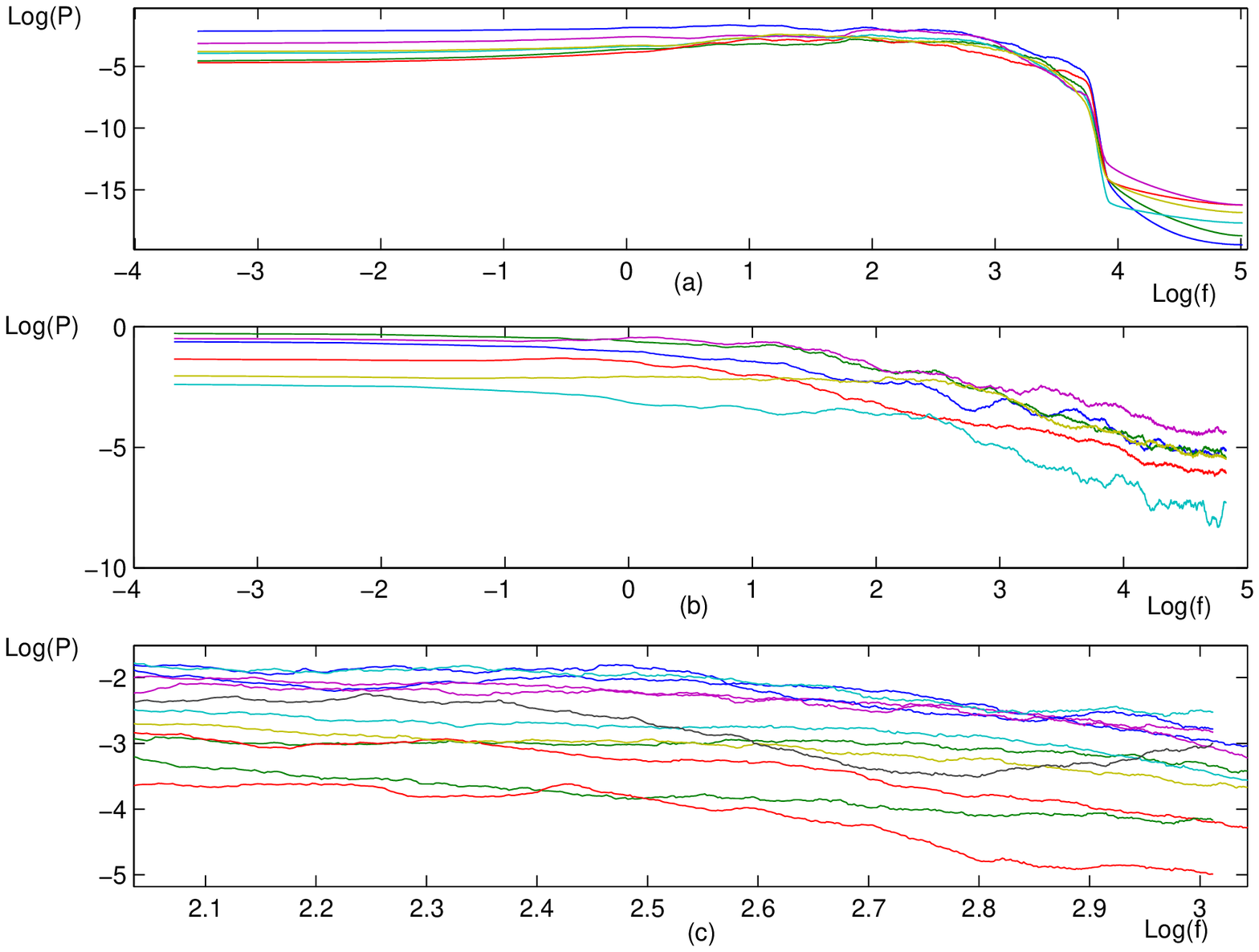}
\end{center}
{\scriptsize Fig. 1 Power spectrum of the ECGs. a) and b) correspond to the whole BBPS of six characteristic pathologic and six characteristic healthy cases respectively. c) corresponds to the same twelve BBPS in the frequency range $\sim 7$Hz to $\sim 20$Hz.}
\end{figure}

These two measures are obtained by analyzing the stochastic information contained in the ECG in the context of Markov processes. First, we must give evidences of Markov properties of the stochastic information. The following equation, Chapman-Kolmogorov, must hold if the time series under analysis corresponds to Markov processes \cite{Ris}:

\begin{equation}
p(x_{1},\tau_{1}\mid x_{3},\tau_{3})=\int dx_{2} p(x_{2},\tau_{2}\mid x_{3},\tau_{3})p(x_{1},\tau_{1}\mid x_{2},\tau_{2})\,
\end{equation}

where $p(x_{1},\tau_{1}\mid x_{2},\tau_{2})$ is the conditional probability density of finding the value $x_{1}$ at time $\tau_{1}$ given the value $x_{2}$ at time $\tau_{2}$. Evidence of such condition may be calculated from the time series and observed graphically in the contour lines of Fig. 2. Similarity of the left plot with the right one is evidence of Markov processes.

\begin{figure}
\begin{center}
\includegraphics[height=8cm,width=9cm]{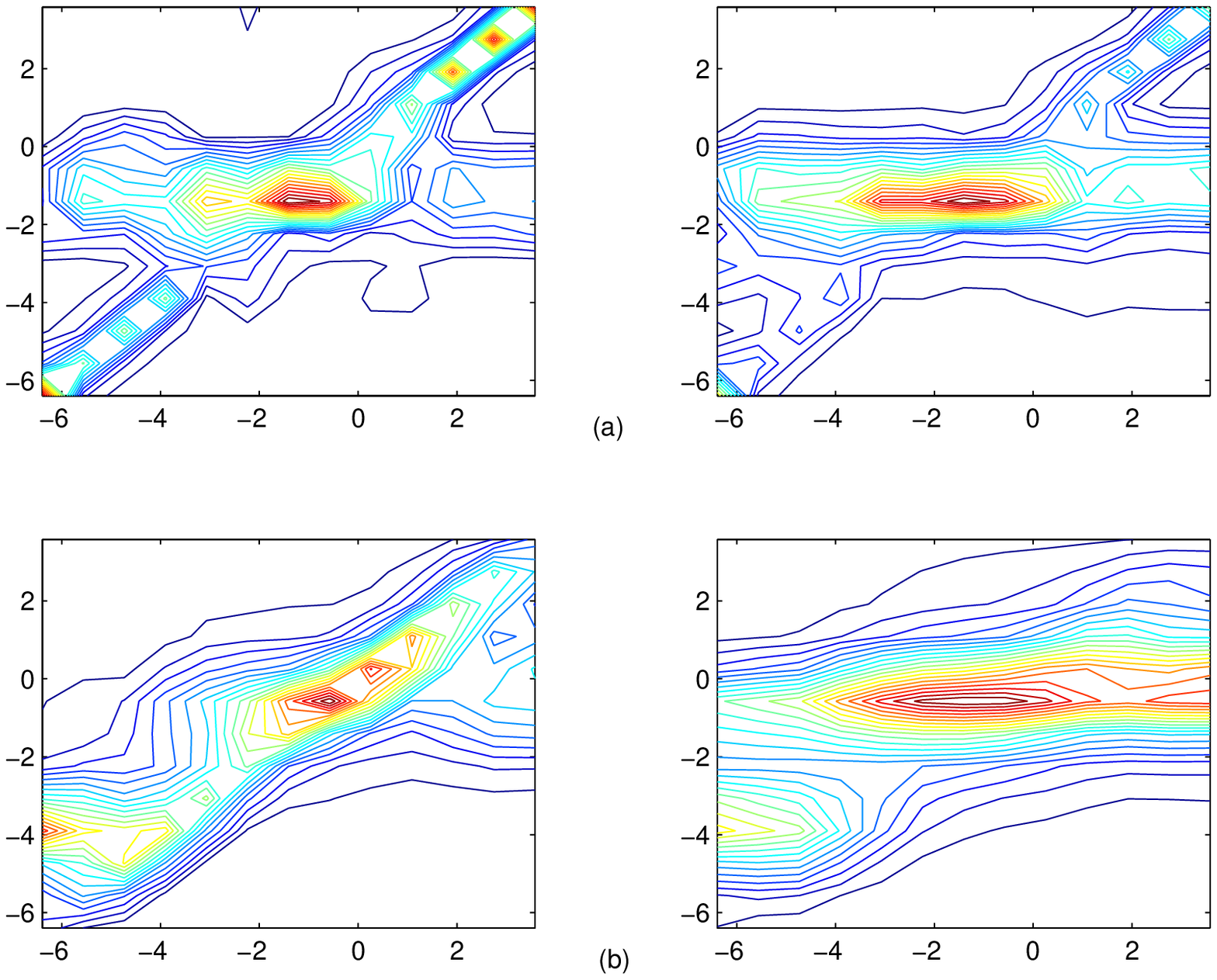}
\end{center}
{\scriptsize Figure 2 Contour plots of the conditional probability distributions $p(x_{1},\tau_{1}\mid x_{3},\tau_{3})$ a) for the healthy case and b) for the pathologic case.}
\end{figure}

The evidence of Markov process corresponding to the SI obtained from all the ECGs may be improved for certain values of the parameters of the method mentioned in section 2, compared with the archetypal reconstruction used in previous works to obtain the SI. With such evidences we can apply an empirical determination of the coefficients $D_{1}$ and $D_{2}$ of the Fokker-Planck equation \cite{RPF}

\begin{equation}
-\tau\frac{\partial}{\partial \tau}p(x,\tau)=\left\{-\frac{\partial}{\partial x}D_{1}(x,\tau)+\frac{\partial^{2}}{\partial x^{2}}D_{2}(x,\tau)\right\} p(x,\tau)\,.
\end{equation}

\noindent
through the Kramers-Moyal coefficients 

\begin{equation}
M_{k}(x,\tau,\Delta\tau)=\frac{\tau}{k!\Delta\tau}\int_{-\infty}^{\infty}(\widetilde{x}-x)^{k}p(\widetilde{x},\tau-\Delta\tau\mid x,\tau)d\widetilde{x}\,,
\end{equation}

\noindent
where {\small $D_{k}(x,\tau)=lim_{\Delta\tau\to 0}M_{k}(x,\tau,\Delta\tau)$}.

The Fokker-Planck equation (2), defines the evolution of the probability distributions $p(x,\tau)$ of the Markov process at time scales defined by $\tau$. The coefficients $D_{1}$ and $D_{2}$, drift and diffusion coefficients respectively, completely determine the Fokker-Planck equation \cite{RPF,Ris,Rei}. Therefore, the coefficients  $D_{1}$ and $D_{2}$ can be used to characterize the stochastic information detected in the ECG as a Markov process.

In figure 3 we present the estimates of the coefficients $D_{1}$ and $D_{2}$ for the SI obtained from one representative pathologic ECG and one representative healthy ECG. We expect that the improvement of the procedure to obtain the SI and the quantification of these results will give a better and more roboust measure to distinguish healthy and pathologic ECG. However, in this work we report the potential of this measure and the method to obtain its values using the ECG as the unique source of information.

\begin{figure}
\begin{center}
\includegraphics[height=7cm,width=8cm]{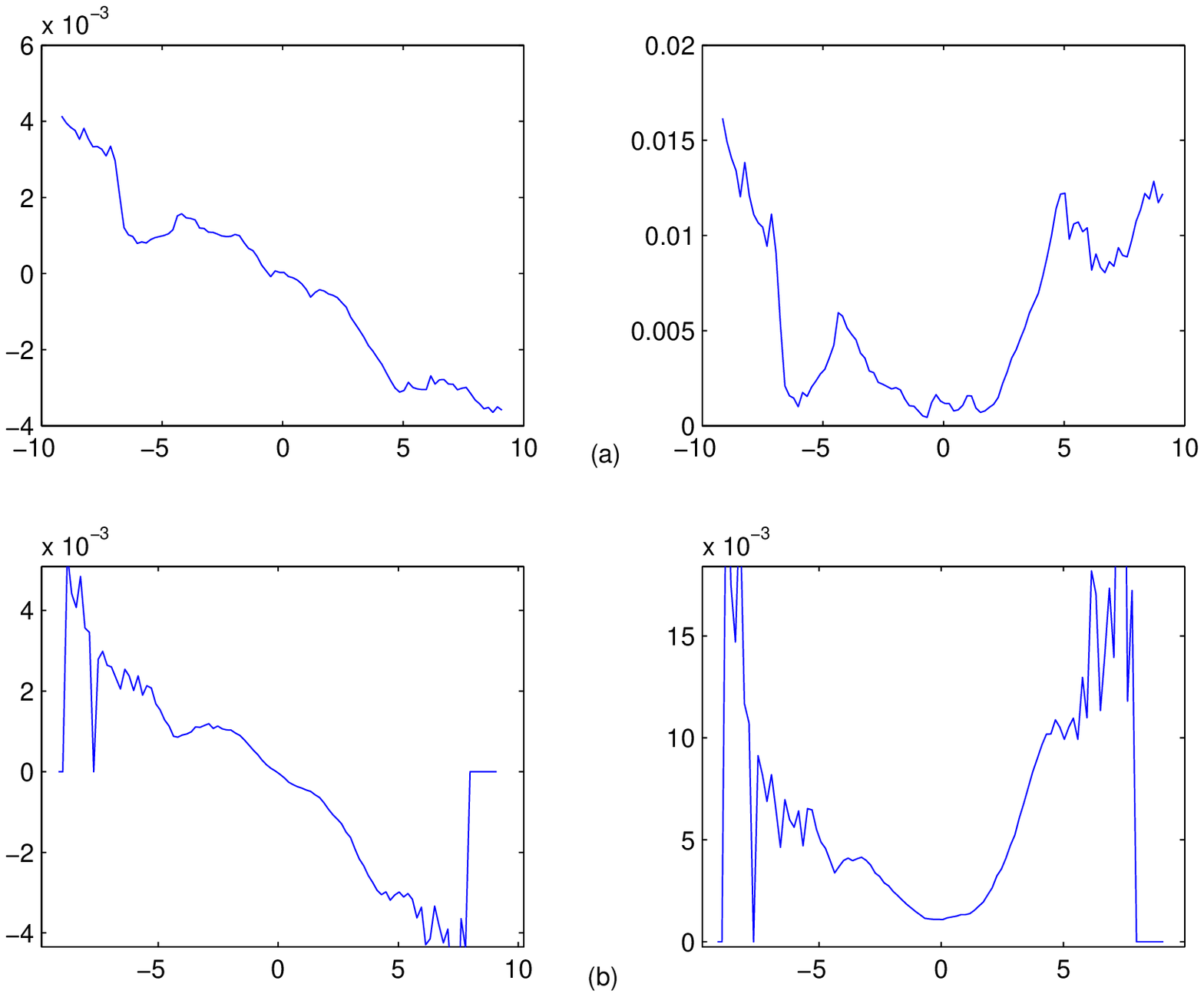}
\end{center}
{\scriptsize Figure 3 Fokker-Planck coefficients,
$D_{1}$ and $D_{2}$ for the stochastic time series obtained from healthy a,b) and pathologic
cases c,d).}
\end{figure}

\section{Discussion and Conclusions}

Despite the preliminary character of the SI obtained from ECGs, the estimated values of the diffusion coefficient $D_{2}$ allows as to identify interesting new differences between healthy and pathologic cases. As it is observed in Fig. 3, the plot of $D_{2}$ as a function of $x$ for the healthy cases did not present a global minimum for $x=0$ and a regular parabolic behavior as was the case for the pathologic cases. This observation is consistent with previous results and a general consensus that healthier cardiac activity corresponds to more complex and less regular behaviors, patterns, measures, etc.
There were two excceptions to this observation, one pathologic and one healthy case. So far, we have not observed relevant information in the different estiamted values of the drift coefficient $D_{1}$, that may be useful to distinguish healthy and pathologic ECGs. 
For all the ECGs analyzed, the corresponding BBPS can be qualitatively classified in two distinctive groups corresponding more or less to the healthy and pathologic ECGs. However, the differences used to define these two groups correspond to very low and very high frequency ranges of the BBPS where external factors of the measuring and recording processes strongly affect the information contained in the ECG. In the frequency range between $\sim 7$Hz and $\sim 20$Hz, we observe that all the BBPS are very similar qualitatively but very different quantitatively as indicated by the values of the measure $\alpha$. However, the different values of $\alpha$ do not allow to make a clear clasification of healthy and pathologic cases.


\begin{thebibliography}{999}


\bibitem{SCI} R.M. Guti\'errez and L. A. Sandoval, Detecting the stochastic and deterministic information of ECGs, proceedings of The $6^{th}$ World Multiconference on Systemics, Cybernetics and Informatics, Orlando-USA, July (2002).


\bibitem{CB} A. Cutler and l. Breiman, Archetypal Analysis, TECHNOMETRICS, V 36, NO. 4, p. 338, November (1994).

\bibitem{Orti} M. D. Ortigueira et al., "An archetypal based ECG analysis system". Downloaded from the http://www.uninova.pt/~mdo/publ.htm. Contact: mdo@uninova.pt. 

\bibitem{http} http://www.physionet.org/ for the 30 pathologic ECGs choosed randomly. The 20 healthy ECGs are from an ECG data base of the Centro de Investigaciones-Universidad Antonio Nariño, Bogotá-Colombia, contact R. M. Guti\'{e}rrez by e-mail: rgutier@uan.edu.co


\bibitem{Ris} H. Risken, The Fokker-Planck Equation, Springer, Berlin, 1984.

\bibitem{Rei} L.E. Reichl, A Moderrn Course in Statistical Physics, University Texas Press, Austin-USA, (1980) Chap. 6. 

\bibitem{RPF} Ch. Renner, J. Peinke and R Friedrich, "Evidence of Markov properties of high frequency exchange rate data", PHYSICA A 298, p. 499 (2001). 


\end{thebibliography}
\end{document}